\documentclass[10pt,conference]{IEEEtran}
\usepackage{array}
\usepackage{cite}
\usepackage{calc}
\usepackage{caption}
\usepackage{graphicx}
\usepackage{psfrag}
\usepackage[tight,footnotesize]{subfigure}
\usepackage{stfloats}
\usepackage{url}
\usepackage{subfig}
\usepackage{longtable}
\usepackage{stfloats}
\usepackage{amsfonts,amssymb,amsbsy,bm,paralist,theorem,cite,ifthen,color}
\usepackage[cmex10]{amsmath}
\usepackage{algorithmic,algorithm}

\newcommand\wb{\ensuremath{{\bm w}}}

\newcommand\hb{\ensuremath{{\bf h}}}
\newcommand\Ab{\ensuremath{{\bf A}}}

\newcommand\Cb{\ensuremath{{\bf C}}}

\newcommand\Ib{\ensuremath{{\bf I}}}

\newcommand\Wb{\ensuremath{{\bf W}}}

\newcommand\zerob{\ensuremath{{\bf 0}}}

\newcommand\E{\ensuremath{{\rm E}}}

\newcommand\tr{\ensuremath{{\rm Tr}}}

\newcommand\Cplx{\ensuremath{{\mathbb{C}}}}

\newcommand{\be}{\ensuremath{{\mathbf{e}}}}
\newcommand{\bh}{\ensuremath{{\mathbf{h}}}}

\newcommand{\bPhi}{\ensuremath{\bm{\Phi}}}
\newcommand{\bPsi}{\ensuremath{\bm{\Psi}}}

\newcommand{\bbh}{\ensuremath{{\bar{\mathbf{h}}}}}

\newcommand{\SINR}{\ensuremath{{\mathrm{SINR}}}}
\newcommand{\st}{\ensuremath{{\mathrm{s.t.}}}}
\newcommand{\Tr}{\ensuremath{{\mathrm{Tr}}}}

\newtheorem{Lemma}{Lemma}

\begin{document}
\bibliographystyle{IEEEtran}
\title{Worst-Case SINR Constrained Robust Coordinated Beamforming for Multicell Wireless Systems}

\author{Chao Shen$^\star$, Kun-Yu Wang$^\dag$, Tsung-Hui
Chang$^{\dag}$, Zhengding Qiu$^{\star}$,
and Chong-Yung Chi$^\dag$ \\ ~ \\
\begin{tabular}{cc}
$^\star$Institute of Information Science                    & $^\dag$Institute of Commun. Eng. \& Department of Elect. Eng. \\
Beijing Jiaotong University,                            & National Tsing Hua University,      \\
Beijing, China, 100044                                  & Hsinchu, Taiwan 30013       \\
\small E-mail: wwellday@gmail.com,~ zdqiu@bjtu.edu.cn   & \small
E-mail: \{kunyuwang7,~tsunghui.chang\}@gmail.com,~
cychi@ee.nthu.edu.tw
\end{tabular}
\vspace{-0.07cm}}

\maketitle
\begin{abstract}
Multicell coordinated beamforming (MCBF) has been recognized as a promising approach to enhancing the system throughput
and spectrum efficiency of wireless cellular systems. In contrast to the conventional single-cell beamforming (SBF)
design, MCBF jointly optimizes the beamforming vectors of cooperative base stations (BSs) (via a central processing
unit (CPU)) in order to mitigate the intercell interference. While most of the existing designs assume that the CPU has
the perfect knowledge of the channel state information (CSI) of mobile stations (MSs), this paper takes into account
the inevitable CSI errors at the CPU, and study the robust MCBF design problem. Specifically, we consider the
worst-case robust design formulation that minimizes the weighted sum transmission power of BSs subject to worst-case
signal-to-interference-plus-noise ratio (SINR) constraints on MSs. The associated optimization problem is challenging
because it involves infinitely many nonconvex SINR constraints. In this paper, we show that the worst-case SINR
constraints can be reformulated as linear matrix inequalities, and the approximation method known as semidefinite
relation can be used to efficiently handle the worst-case robust MCBF problem. Simulation results show that the
proposed robust MCBF design can provide guaranteed SINR performance for the MSs and outperforms the robust SBF design.
\end{abstract}

\IEEEpeerreviewmaketitle

\section{Introduction}
Recently, multicell cooperative signal processing has drawn considerable attention since it, when
compared with the conventional single-cell processing, can provide significant system throughput
gains by exploiting the degrees of freedom provided by multiple multi-antenna base stations (BSs).
In contrast to the single-cell transmission design which treats the interference from neighboring
cells as noise, in the multicell cooperative system, BSs collaborate with each other to jointly
design their transmissions in order to mitigate the intercell interference
\cite{Luca2010TWC,Dahrouj2010,Kim2009,Huh2010,Zhangrui2010}. This paper considers the multicell
coordinated beamforming (MCBF) design \cite{Dahrouj2010,Kim2009} where a set of multiple-antenna
BSs jointly design their beamforming vectors aiming at providing desired quality-of-service (QoS)
for the mobile stations (MSs). To this end, it is assumed that the BSs are connected with a central
processing unit (CPU) (which can be a dedicated control center or a preselected BS), which knows
all the channel state information (CSI) of MSs. With the perfect CSI, it has been shown that the
MCBF design problem can be efficiently solved via convex
optimization theory \cite{Dahrouj2010}. 

In practical systems, however, the CSI available to the CPU may not be perfect. In particular, the CSI may be subject
to channel estimation errors due to finite-length training, and quantization errors owing to limited feedback bandwidth
(of the channels from the MSs to BSs). The imperfect CSI may result in performance outage and the QoS requirements of
MSs can no longer be guaranteed. In view of this, transmit beamforming designs that take the CSI errors into
consideration, also known as robust transmit beamforming, are of great importance to maintain the QoS of MSs.

In this paper, we assume elliptically bounded CSI errors, and study the robust MCBF design problem. Specifically, we
consider the worst-case robust design formulation that minimizes the weighted sum transmission powers of the BSs
subject to worst-case signal-to-interference-plus-noise ratio (SINR) constraints on the MSs. This robust formulation
guarantees the MSs to achieve the desired SINR performance for all possible CSI errors. The worst-case robust design
formulation has been studied in the context of single-cell robust transmit beamforming; see
\cite{Shenouda07JSTSP,Shenouda2008JSAC,ShenouaAsilomar2008,Vucic_Boche_2009_MIMO,Wang2010}. However, the robust
formulation for MCBF is more challenging since the associated SINR constraints involve CSI errors not only in the
desired signal and intra-cell interference terms, but also in the intercell interference. In this paper, we show that
the worst-case robust MCBF problem can be efficiently handled by semidefinite relaxation (SDR), a convex optimization
based approximation method \cite{Luo2010_SPM}. Specifically, it can be shown that the worst-case SINR constraints can
be recast as a finite number of linear matrix inequalities (LMIs), and SDR can be applied to approximate the original
nonconvex problem by a convex semidefinite program (SDP), which, thereby, can be efficiently solved \cite{BK:BoydV04}.
The presented simulation results show that the proposed worst-case robust MCBF design can provide guaranteed SINR
performance for the MSs, and is more power efficient and more feasible than the conventional single-cell robust
beamforming design.

\vspace{-0.1cm}
\section{System Model And Problem Statement}

We consider a multicell wireless system with $N_c$ cells. Each cell consists of a BS, which is equipped with $N_t$
antennas, and $K$ single-antenna MSs; see Fig. 1 for an example of $N_c=3$ and $K=4$. The $N_c$ BSs will collaborate to
enhance the strength of the signal of interest for each MS while mitigating the intercell interference. Let $s_{ik}(t)$
be the information signal for MS $k$ in the $i$th cell with $\E\{|s_{ik}(t)|^2\}\!=\!1$; and let $\wb_{ik}\in
\Cplx^{N_t}$ be the associated beamforming vector, $\{\wb_{ik}\}$ be the set of all beamforming vectors, i.e.,
$\{\wb_{ik}\}\triangleq\{\wb_{11},\ldots,\wb_{N_cK}\}$. The transmit signal by the $i$th BS is given by
$\sum_{k=1}^K{\wb_{ik}s_{ik}(t)}$ for $i=1,\ldots,N_c$. Denote by $\hb_{jik}\in \mathbb{C}^{N_t}$ the channel vector
from the $j$th BS to the $k$th MS in the $i$th cell and denote by $\{\hb_{jik}\}_{j=1}^{N_c}$ the set of channel
vectors from all BSs to the $k$th MS in cell $i$. The received signal of MS $k$ in the $i$th cell can be expressed as
\begin{subequations}\label{RcvSig}
\begin{align}
\!\!\!\!y_{ik}(t) =&\sum_{j=1}^{N_c}\bh_{jik}^H\left(\sum_{\ell=1}^K{\wb_{j\ell}s_{j\ell}(t)}\right)+z_{ik}(t)\\
    =& \hb_{iik}^H\wb_{ik}s_{ik}(t)+\sum_{\substack{{\ell\neq k}}}^K{\hb_{iik}^H\wb_{i\ell}s_{i\ell}(t)}\notag\\ 
    &~~~~~~~~~~~~~~~~+\sum_{\substack{{j\neq i}}}^{N_c}%
    {\hb_{jik}^H{\sum_{\ell=1}^{K}\wb_{j\ell}s_{j\ell}(t)}}
    +z_{ik}(t),\label{RcvSig-2}
\end{align}
\end{subequations}
where the first term in \eqref{RcvSig-2} is the signal of interest, the second and third terms are
the intra-cell and intercell interference, respectively, and $z_{ik}(t)$ is the additive noise with
zero mean and variance $\sigma_{ik}^2>0$. From \eqref{RcvSig}, the SINR of the $k$th MS in the
$i$th cell can be shown to be
\begin{align}\label{SINR}
&{{\SINR}_{ik}}\left(\{\wb_{j\ell}\},\{\hb_{jik}\}_{j=1}^{N_c}\right)  \notag \\
&~~~~~~~~~~~=\frac{{\left|\hb_{iik}^H\wb_{ik}\right|}^2}%
{\sum\limits_{\ell\neq k}^K {\left|\hb_{iik}^H\wb_{i\ell}\right|}^2%
+\sum\limits_{j\neq i}^{N_c}\sum\limits_{\ell=1}^K {\left|\hb_{jik}^H\wb_{j\ell}\right|}^2%
+\sigma_{ik}^2}.%
\end{align}

Using the SINR in \eqref{SINR} as the MSs' QoS measure and under the assumption that the CPU has the perfect knowledge
of all the channels $\{\hb_{jik}\}$, the following design formulation
\begin{subequations}\label{PM}
\begin{align}
  \min_{\{\wb_{ik}\}}~&\sum_{i=1}^{{N_c}}
  {\alpha _i}\left({\sum_{k = 1}^K
  {\left\| \wb_{ik}\right\|^2} }\right)\\
  \st~~&
  {{\SINR}_{ik}}\left(\{\wb_{j\ell}\},\{\hb_{jik}\}_{j=1}^{N_c}\right) \!\geq\!\gamma_{ik}, \notag
  \\&k=1,\ldots,K,~i=1,\ldots,N_c,
\end{align}
\end{subequations}
has been considered in \cite{Dahrouj2010}, where $\alpha_i>0$ is the power weight for BS $i$, and $\gamma_{ik}>0$ is
the target SINR for MS $k$ in cell $i$. One can see from \eqref{PM} that the $N_c$ BSs jointly design their beamforming
vectors such that the weighted sum power of BSs is minimized while each of the MSs can achieve the desired SINR
specification $\gamma_{ik}$. It has been shown that problem \eqref{PM} can be reformulated as a convex second-order
cone program (SOCP) and can be efficiently solved via standard solvers, e.g., \texttt{CVX} \cite{cvx}.

In addition to problem \eqref{PM}, we also consider the conventional single-cell beamforming (SBF) design that avoids
interfering neighboring cells by per-cell interference control \cite{Huh2010}, i.e., BS $i$ designs the beamforming
vectors $\{\wb_{ik}\}_{k=1}^K$ independently by solving the following problem:
\begin{align}\label{static PM}
\!\!\!\min_{\substack{\{\wb_{ik}\}_{k=1}^K}}~&{\sum_{k = 1}^K
  {\left\| \wb_{ik}\right\|^2} }\\
  \st~~&
  \frac{{\left|\hb_{iik}^H\wb_{ik}\right|}^2}%
{\sum\limits_{\ell\neq k}^K {\left|\hb_{iik}^H\wb_{i\ell}\right|}^2%
+\sum\limits_{j\neq i}^{N_c} \xi_{jik}%
+\sigma_{ik}^2}\!\geq\!\gamma_{ik}, k=1,\ldots,K, \notag \\
 &\sum_{k=1}^K {\left|\hb_{ij\ell}^H\wb_{ik}\right|}^2 \leq
  \xi_{ij\ell},~\ell=1,\ldots,K,~j\in \mathcal{N}_c\backslash\{i\}, \notag
\end{align}
for $i=1,\ldots,N_c$, where $\mathcal{N}_c=\{1,\ldots,N_c\}$, and $\xi_{ij\ell}>0$ stands for the
preset, maximum tolerable interference from BS $i$ to the $\ell$th user in cell $j$. As seen from
\eqref{static PM}, the SBF design conservatively treats the intercell interference upper bound
$\xi_{jik}$ as fixed noise powers, in contrast to the MCBF design in \eqref{PM} where the $N_c$ BSs
collaborate to dynamically control the intercell interference. It has been shown that the SBF
design is less power efficient than the MCBF design in \eqref{PM} \cite{Dahrouj2010}; however the
SBF design can inherently be implemented at each BS in a decentralized fashion.

\begin{figure}[t]
\centering
\includegraphics[width=5cm]{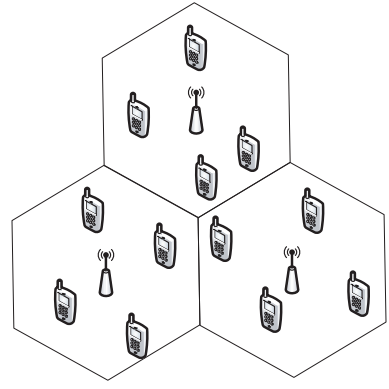}\vspace{0.3cm}
\caption{An example of wireless cellular system with 3 BSs and 4 MSs
in each cell.} \label{fig:figure1} \vspace{-0.3cm}
\end{figure}

\section{Proposed Robust Coordinated Beamforming}

\subsection{Robust MCBF}
The above formulations in \eqref{PM} and \eqref{static PM} assume that the CPU knows the exact CSI
$\{\hb_{jik}\}$. In the case that the CPU has CSI with errors, the standard formulations in
\eqref{PM} and \eqref{static PM} can no longer guarantee the desired SINR requirements. To resolve
this problem, we consider the worst-case robust design formulation
\cite{Shenouda07JSTSP,Shenouda2008JSAC}.

Specifically, we model the true channel vector $\hb_{jik}$ as
\begin{equation}\label{CSIErrorExp}
{\hb_{jik}} = \bbh_{jik} + \be_{jik},
\end{equation} for $k=1,\ldots,K$, $i,j\in\mathcal{N}_c$,
where $\be_{jik}\!\in\!\mathbb{C}^{N_t}$ represents the channel error vector. Moreover, let us consider the
elliptically bounded CSI errors, that is, each $\be_{jik}$ satisfies
\begin{align}\label{CSIErrorExp2}
\be_{jik}^H\Cb_{jik}\be_{jik}\leq 1,
\end{align}
where $\Cb_{jik}\succ 0$ (a positive definite matrix) determines the size and the shape of the
error ellipsoid. With \eqref{CSIErrorExp} and \eqref{CSIErrorExp2}, we consider the following
worst-case SINR constraint on MS $k$ in cell $i$:
\begin{align}\label{worst case SINR constraint}
  &{{\SINR}_{ik}}\left(\{\wb_{j\ell}\},\{\bbh_{jik}+\be_{jik}\}_{j=1}^{N_c}\right) \!\geq\!\gamma_{ik}~ \notag \\
  &~~~~~~~~~~~~~~~~~~~~~~\forall~ \be_{jik}^H\Cb_{jik}\be_{jik}\leq
  1,~j=1,\ldots,N_c.
\end{align}
Note from \eqref{worst case SINR constraint} that the SINR specification $\gamma_{ik}$ is satisfied for all possible
CSI errors. Taking the worst-case SINR constraints in \eqref{worst case SINR constraint} into consideration, we obtain
the following design formulation
\begin{subequations}\label{RPM}
\begin{align}
  \min_{\substack{\{\wb_{ik}\}}}~&\sum_{i=1}^{{N_c}}
  {\alpha _i}\left({\sum_{k = 1}^K
  {\left\| \wb_{ik}\right\|^2} }\right)\\
  \st~~&
  {{\SINR}_{ik}}\left(\{\wb_{j\ell}\},\{\bbh_{jik}+\be_{jik}\}_{j=1}^{N_c}\right) \!\geq\!\gamma_{ik}~ \notag \\
  &\forall~ \be_{jik}^H\Cb_{jik}\be_{jik}\leq
  1,~j=1,\ldots,N_c, \label{RPM-C1} \\
  &i=1,\ldots,N_c,~k=1,\ldots,K,\notag
\end{align}\end{subequations}
as a worst-case robust counterpart of problem \eqref{PM}. Solving the optimization problem
\eqref{RPM} is challenging due to the infinitely many nonconvex SINR constraints in \eqref{RPM-C1}.
To handle this problem, let us present a suboptimal method via SDR and S-procedure
\cite{BK:BoydV04} in the next subsection.

\subsection{Solving \eqref{RPM} by SDR and S-Procedure }

Let us express the objective function of problem \eqref{RPM} as $\sum_{i=1}^{{N_c}}{\alpha
_i}{\sum_{k = 1}^K {\tr(\wb_{ik}\wb_{ik}^H)} }$, where $\tr(\cdot)$ denotes the trace of a matrix,
and express the worst-case SINR constraint of the $k$th MS in the $i$th cell [in \eqref{worst case
SINR constraint}] as
\begin{align}\label{SINR_constraint}
&\!\!{\left(\bbh_{iik}^H+\be_{iik}^H\right)\left(\frac{1}{\gamma_{ik}}%
\wb_{ik}\wb_{ik}^H-\sum_{\ell\neq k}^K \wb_{i\ell}\wb_{i\ell}^H\right)\left(\bbh_{iik}+\be_{iik}\right)}\notag\\
&\geq\sum_{j\neq
i}^{N_c}{\left(\bbh_{jik}^H+\be_{jik}^H\right)\left(\sum\limits_{\ell=1}^K\wb_{j\ell}\wb_{j\ell}^H\right)\left(\bbh_{jik}+\be_{jik}\right)}
+\sigma_{ik}^2\notag \\
&\forall~ \be_{jik}^H\Cb_{jik}\be_{jik}\leq 1,~j=1,\ldots,N_c.
\end{align}

The idea of SDR is to replace each rank-one matrix $\wb_{ik}\wb_{ik}^H$ by a general-rank positive
semidefinite matrix $\Wb_{ik}$, i.e., $\Wb_{ik} \succeq \zerob$ \cite{Luo2010_SPM}. By applying SDR
to \eqref{RPM}, we obtain the following problem
\begin{subequations}\label{RPM SDR}
\begin{align}
  \min_{\substack{\{\Wb_{ik}\succeq \zerob\}}}&\sum_{i=1}^{{N_c}}
  {\alpha _i}{\left(\sum_{k = 1}^K{\tr(\Wb_{ik})}\right)}\\
  \st~&
  {\left(\bbh_{iik}^H+\be_{iik}^H\right)\!\!\left(\!\frac{1}{\gamma_{ik}}
    \Wb_{ik}\!-\!\sum_{\ell\neq k}^K \Wb_{i\ell}\!\right)\!\!
    \left(\bbh_{iik}+\be_{iik}\right)}\notag\\
    &\geq\!\sum_{j\neq
    i}^{N_c}{\left(\bbh_{jik}^H+\be_{jik}^H\right)\!\!\left(
    \!\sum\limits_{\ell=1}^K\Wb_{j\ell}\right)\!\!\left(\bbh_{jik}+\be_{jik}\right)}
    \!+\!\sigma_{ik}^2, \notag \\
  &~\forall~ \be_{jik}^H\Cb_{jik}\be_{jik}\leq
  1,~j=1,\ldots,N_c, \label{RPM SDR-C2}\\
  &~i=1,\ldots,N_c,~k=1,\ldots,K.\notag
\end{align}\end{subequations}
While the SDR problem \eqref{RPM SDR} is convex, it is still difficult to handle owing to an
infinite number of linear inequality constraints. To resolve this, we observe that the left-hand
side and right-hand side of the first inequality in \eqref{RPM SDR-C2} involve independent CSI
errors. Hence, the constraint \eqref{RPM SDR-C2} can be equivalently decoupled into the following
$N_c$ constraints:
\begin{align}\label{SINR_constraint2-1}
&{\left(\bbh_{iik}^H+\be_{iik}^H\right)\left(\frac{1}{\gamma_{ik}} \Wb_{ik}-\sum_{\ell\neq k}^K \Wb_{i\ell}\right)\left(\bbh_{iik}+\be_{iik}\right)}\notag\\
&~~~~~~~~~~~~~~~~~~\geq\sum_{j\neq i}^{N_c}t_{jik} +\sigma_{ik}^2 ~\forall~\be_{iik}^H\Cb_{iik}\be_{iik}\leq 1, \\
&{\left(\bbh_{jik}^H+\be_{jik}^H\right)
\left(\sum\limits_{\ell=1}^K\Wb_{j\ell}\right)\left(\bbh_{jik}+\be_{jik}\right)}\leq t_{jik}~\notag \\
&~~~~~~~~~~~~~~~~~~~~~~\forall~ \be_{jik}^H\Cb_{jik}\be_{jik}\leq 1,~j\in \mathcal{N}_c\backslash
\{i\}, \label{SINR_constraint2-2}
\end{align}
where $\{t_{jik}\}_{j\neq i}$ are slack variables. Note that equation \eqref{SINR_constraint2-1}
involves only the CSI error $\be_{iik}$ and each of the constraints in \eqref{SINR_constraint2-2}
involves only one CSI error $\be_{jik}$. Furthermore, \eqref{SINR_constraint2-1} and
\eqref{SINR_constraint2-2} can be reformulated as finite LMIs, by applying the following
S-procedure:
\begin{Lemma}\cite[S-procedure]{BK:BoydV04}\label{sprocedure}
Let $\Ab{},\Cb\in\mathbb{C}^{N_t\times N_t}$ be complex Hermitian matrices,
${\be}\in\mathbb{C}^{N_t}$ and $c\in\mathbb{R}$. The following condition
\begin{equation*} \be^H\Ab{}\be+{\bf{b}}^H\be+\be^H {\bf{b}}+c\geq 0~~\forall~\be^H\Cb\be\leq 1\end{equation*}
holds true if and only if there exists a $\lambda \geq 0$ such that%
\begin{align*}
\begin{bmatrix}\Ab{}+\lambda\Cb & {\bf{b}}\\ {\bf{b}}^H & c-\lambda\end{bmatrix}\succeq \zerob.
\end{align*}
\end{Lemma}
By applying Lemma 1, one can recast \eqref{SINR_constraint2-1} as
\begin{align}\label{LMI1}
&\bPhi_{ik}\left(\{\Wb_{i\ell}\}_{\ell=1}^K,\{t_{jik}\}_{j\neq i},\lambda_{iik}\right) \triangleq
\notag \\
&~~~~~~~\begin{bmatrix}
  \Ib \\
   \bar\hb_{iik}^H
   \end{bmatrix}\left(\frac{1}{\gamma_{ik}}
\Wb_{ik}-\sum_{\ell\neq k}^K \Wb_{i\ell}\right)
\begin{bmatrix}
  \Ib \\
   \bar\hb_{iik}^H
   \end{bmatrix}^H\notag \\
    &~~~~~~~~~~~~~+\begin{bmatrix}
    \lambda_{iik}\Cb_{iik} & \zerob \\
                 \zerob    & -\sum\limits_{j\neq i}^{N_c}t_{jik}\!\!-\!\sigma_{ik}^2\!\!-\!\!\lambda_{iik}
   \end{bmatrix}\succeq \zerob,
\end{align}
where $\Ib$ is the $N_t\times N_t$ identity matrix, and recast
\eqref{SINR_constraint2-2}, for each $j\in
\mathcal{N}_c\backslash\{i\}$, as
\begin{align}\label{LMI2}
  &   \bPsi_{jik}\left(\{\Wb_{j\ell}\}_{\ell=1}^K,t_{jik},\lambda_{jik}\right)\triangleq
\begin{bmatrix}
  \Ib \\
   \bar\hb_{jik}^H
   \end{bmatrix}\!\!\left(\!-\sum_{\ell=1}^K \Wb_{j\ell}\right)\!
\begin{bmatrix}
  \Ib \\
   \bar\hb_{jik}^H
   \end{bmatrix}^H\notag \\
   &+\begin{bmatrix}
    \lambda_{jik}\Cb_{jik} & \zerob \\
                 \zerob    & t_{jik}-\lambda_{jik}
   \end{bmatrix}\succeq \zerob,
\end{align}
where $\lambda_{jik}\geq 0$ for all $i,j=1,\ldots,N_c,$ and
$k=1,\ldots,K$. 

Replacing \eqref{RPM SDR-C2} with \eqref{LMI1} and \eqref{LMI2} leads to the following SDR problem
\begin{align}\label{DCBF-SDP}
\min_{\substack{\{\Wb_{ik}\},\{\lambda_{jik}\},\\\{t_{jik},j\neq i\}}}~
  &\sum_{i=1}^{N_c}\alpha_i\left( \sum_{k=1}^K\Tr(\Wb_{ik})\right)\\
  \st~
  &\bPhi_{ik}\left(\{\Wb_{i\ell}\}_{\ell=1}^K,\{t_{jik}\}_{j\neq i},\lambda_{iik}\right)
  \succeq \zerob, \notag \\
  &\bPsi_{jik}\left(\{\Wb_{j\ell}\}_{\ell=1}^K,t_{jik},\lambda_{jik}\right)
  \succeq \zerob,~ j\in \mathcal{N}_c\backslash\{i\}, \notag \\
  &t_{jik}\geq 0,~ j\in \mathcal{N}_c\backslash\{i\}, \notag \\
  &\Wb_{ik}\succeq \zerob,~\lambda_{jik}\geq 0,~ j\in \mathcal{N}_c, \notag \\
  &i=1,\ldots,N_c,~k=1,\ldots,K. \notag
\end{align}
Problem \eqref{DCBF-SDP} is a convex semidefinite program (SDP); hence it can be efficiently solved
\cite{cvx}.

Similarly, one can also consider a worst-case robust design for the SBF design in \eqref{static
PM}, which is given by
\begin{align}\label{static RPM}
\!\!\!\min_{\substack{\{\wb_{ik}\}_{k=1}^K}}~&{\sum_{k = 1}^K
  {\left\| \wb_{ik}\right\|^2} }\\
  \st~~&
  \frac{{\left|(\bar\hb_{iik}+\be_{iik})^H\wb_{ik}\right|}^2}%
{\sum\limits_{\ell\neq k}^K {\left|(\bar\hb_{iik}+\be_{iik})^H\wb_{i\ell}\right|}^2%
+\sum\limits_{j\neq i}^{N_c} \xi_{jik}%
+\sigma_{ik}^2}\!\geq\!\gamma_{ik}~\notag\\
&\forall~\be_{iik}^H\Cb_{iik}\be_{iik}\leq 1,~k=1,\ldots,K, \notag \\
&\sum_{k=1}^K {\left|(\bar\hb_{ij\ell}+\be_{ij\ell})^H\wb_{ik}\right|}^2 \leq \xi_{ij\ell}~\forall~\be_{ij\ell}^H\Cb_{ij\ell}\be_{ij\ell}\leq 1, \notag \\%
&\ell=1,\ldots,K,~j\in \mathcal{N}_c\backslash\{i\}, \notag
\end{align}
for $i=1,\ldots,N_c$. By using similar techniques of S-procedure and SDR, one can obtain an SDR
problem for \eqref{static RPM} which can also be efficiently handled.

Since the SDR problem \eqref{DCBF-SDP} is obtained by rank relaxation of problem \eqref{PM}, the
obtained optimal $\{\Wb_{ik}\}$ of \eqref{DCBF-SDP} may not be of rank one. If the obtained optimal
$\{\Wb_{ik}\}$ happens to be of rank one, i.e., $\Wb_{ik}=\wb_{ik}\wb_{ik}^H$ for all $i$, $k$,
then $\{\wb_{ik}\}$ is an optimal solution of the original problem \eqref{RPM}; otherwise
additional solution approximation procedure is needed; see \cite{Luo2010_SPM} for the details.
Interestingly, it is observed in our simulations that problem \eqref{DCBF-SDP} always yields
rank-one optimal $\{\Wb_{ik}\}$, which implies that an optimal solution of problem \eqref{RPM} can
always be obtained for the problem instances in our simulations. The same rank-one optimality
results are also observed for the SDR problem of problem \eqref{static RPM}.

\section{Simulation Results and Discussions}
In this section, some simulation results are presented to demonstrate the performance of the proposed robust MCBF
design. We consider a multicell system with three cells ($N_c=3$) and two MSs ($K=2$) in each cell. Assume that each BS
has five antennas ($N_t=5$) and the inter-BS distance is 500 meters. In the simulations, we incorporate both
large-scale and small-scale channel fadings. Specifically, we define the true channel $\{\hb_{jik}\}$ with parameters
taken from the 3GPP Long Term Evolution (LTE) channel model \cite{LTESCM}, as follows:
\begin{align}
\hb_{jik}=10^{\frac{34.6+35{\log_{10}}(d_{jik})}{-20}}\cdot\psi_{jik}\cdot\varphi_{jik}\cdot
(\bar\hb_{jik}+\be_{jik}),\label{channel model}
\end{align}
where the exponential term is due to the path loss depending on the distance between the $j$th BS and the $k$th MS in
cell $i$ (denoted by $d_{jik}$ in meters), $\psi_{jik}$ reflects the shadowing effect, $\varphi_{jik}$ represents the
transmit-receive antenna gain, and the term inside the parenthesis denotes the small-scale fading which is composed of
the channel estimate $\bar\hb_{jik}$ and the CSI error $\be_{jik}$. As seen from \eqref{channel model}, it is assumed
that the CPU can accurately track the large-scale fading with CSI errors only in the small-scale fading. In the
simulations, the locations of the two MSs in each cell are randomly determined (with distance to the associated BS at
least 35 meters, i.e., $d_{iik}\geq 35$ for all $i,k$), and thereby the distances to neighboring BSs, i.e.,
$\{d_{jik}\}_{j\neq i}$, can be determined. The shadowing coefficient $\psi_{jik}$ follows the log-normal distribution
with zero mean and standard deviation equal to $8$. The elements of the channel estimate $\{\bar\hb_{jik}\}$ are
independent and identically distributed complex Gaussian random variables with zero mean and unit variance. For
simplicity, we assume the spherically bounded CSI errors, i.e., $\Cb_{jik}=1/\epsilon^2 \Ib$, with the uncertainty
radius $\epsilon$ set to 0.1. We also assume that all the MSs have the same noise power equal to
$\sigma_{jik}^2\triangleq \sigma^2\!=-106.27$ dBm \cite{Dahrouj2010}, the same target SINRs, i.e.,
$\gamma_{jik}\triangleq \gamma$, and the same antenna gains, i.e., $\varphi_{jik}=5$ dBi for all $j$, $i$ and $k$. We
consider the total sum power minimization problem for formulations in \eqref{PM}, \eqref{static PM}, \eqref{RPM} and
\eqref{static RPM} by setting $\alpha_i=1$ for all $i=1,\ldots,N_c$. For problems \eqref{static PM} and \eqref{static
RPM}, we set all the intercell interference constraints $\{\xi_{jik}\}_{j\neq i}$ equal to the noise power $\sigma^2$
\cite{Huh2010}. The robust formulations in \eqref{RPM} and \eqref{static RPM} are handled by the proposed SDR method
described in Sec. III-B, and \texttt{CVX} \cite{cvx} is used to solve the associated SDPs.

In the first example, we investigate the minimal achievable SINRs of the four formulations, namely,
the non-robust SBF design in \eqref{static PM}, the non-robust MCBF design in \eqref{PM} and their
robust counterparts in \eqref{static RPM} and \eqref{RPM}, in the presence of CSI errors. We
generated $5,000$ sets of channel estimates $\{\bar\hb_{jik}\}$, and, for each set of
$\{\bar\hb_{jik}\}$, $100$ sets of CSI errors $\{\be_{jik}\}$ satisfying $\|\be_{jik}\|^2 \leq
\epsilon^2$ were uniformly generated to evaluate the achievable SINRs [in \eqref{SINR}] by the four
formulations. Figure 2 shows the simulation results of the minimal achievable SINR among all the
MSs, by averaging over the channel estimates for which the four formulations under test are all
feasible. It can be observed from this figure that both the robust designs in \eqref{static RPM}
and \eqref{RPM} can guarantee the minimal SINR of MSs no less than the target SINR $\gamma$;
whereas the non-robust designs can have SINR far below $\gamma$ due to the CSI errors. In
particular, one can see from Fig. 2 that, for $\gamma=9$ dB, the minimal SINR achieved by
non-robust MCBF is more than 10 dB lower than that achieved by robust MCBF.

\begin{figure}[t]
\includegraphics[width=0.47\textwidth]{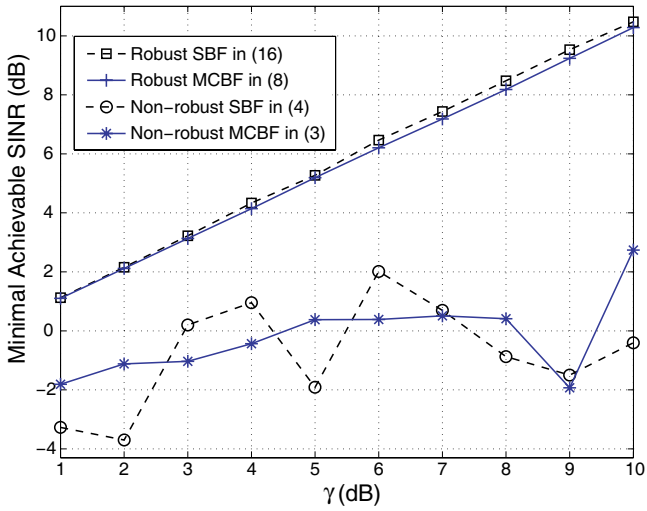}
\caption{Minimal achievable SINR of MSs versus target SINR $\gamma$. } \label{SINRmin}\vspace{0cm}
\end{figure}
\begin{figure}[t]
\includegraphics[width=0.47\textwidth]{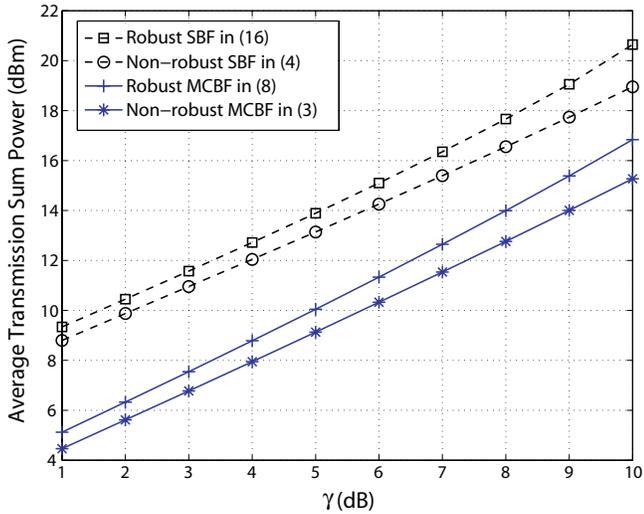}
\caption{Average transmission sum power versus target SINR
$\gamma$.} \label{PowervsSINRtarget}\vspace{-0.3cm}
\end{figure}

To show the power efficiency of MCBF, we present in Fig. 3 the corresponding average transmission powers of the four
methods under test. As a price for worst-case performance guarantee, one can observe from this figure that the robust
designs in \eqref{static RPM} and \eqref{RPM} require more transmission powers than their non-robust counterparts in
\eqref{static PM} and \eqref{PM}, respectively. However, the robust MCBF design ({$+$}) has an average sum power which
is around $4$ dBm less than that of the robust SBF design (${\square}$). Comparing Fig. 3 with Fig. 2, one can see that
the robust MCBF is more power efficient than the robust SBF in achieving the same SINR performance. Finally, we show
the feasibility rates of the four formulations under test in Fig. 4. As seen, the robust designs have lower feasibility
rates compared to their non-robust counterparts; whereas, the proposed robust MCBF design ($+$) exhibits a
significantly higher feasibility rate than the robust SBF design ($\square$) since the former design makes use of the
full degrees of freedom of the multicell system in intercell interference suppression.

\begin{figure}[t]
\vspace{-0.15cm}
\includegraphics[width=0.47\textwidth]{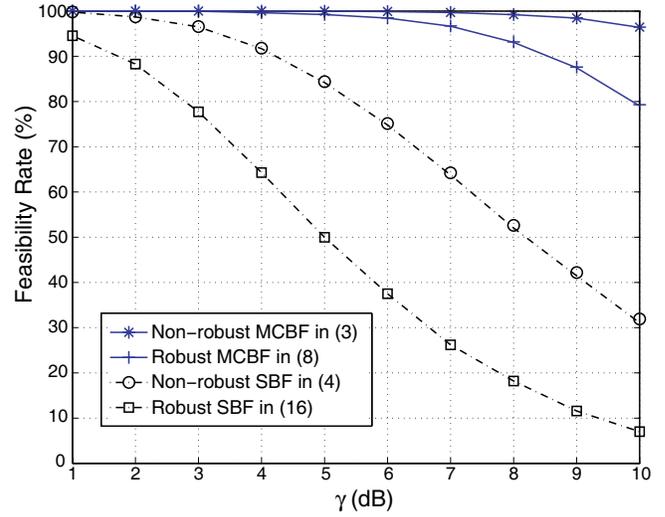} \caption{Feasibility rate (\%)
versus target SINR $\gamma$.} \label{feasratio}\vspace{0.05cm}
\end{figure}

\section*{Acknowledgments}
This work is supported partly by National Science Council (NSC), R.O.C., under grants NSC 98-2219-E-007-005 and NSC
99-2221-E-007-052-MY3. The first author Chao Shen is supported by the Nufront Fellowship.
\vspace{0.26cm}

\bibliography{ICC2011ref}
\end{document}